\documentclass[letterpaper]{JHEP3}
\usepackage{epsfig,psfrag}

\newcommand{\be}{\begin{equation}}
\newcommand{\ee}{\end{equation}}
\newcommand{\bear}{\begin{eqnarray}}
\newcommand{\eear}{\end{eqnarray}}
\newcommand{\ba}{\begin{array}}
\newcommand{\ea}{\end{array}}

\title{Electroweak Symmetry Breaking from a \\
Holographic Fourth Generation
\\ { $\; $ } \\ }

\author{Gustavo Burdman and  Leandro Da Rold\\
Departamento de F\'{i}sica Matem\'{a}tica\\ 
Instituto de F\'{i}sica, Universidade de S\~{a}o Paulo,
\\ R. do Mat\~{a}o 187, S\~{a}o Paulo, SP 05508-900, Brazil \\ 
\\ \\
\email{burdman@if.usp.br, daroldl@fma.if.usp.br} \\ }

\abstract{
We consider a model with
four generations of standard model fermions propagating in an extra dimension 
with an AdS background metric. 
We show that if the zero modes of the fourth generation are  highly
localized towards the infrared brane,  
it is possible to break the electroweak symmetry via 
their condensation, partly  driven by their interactions with the 
Kaluza-Klein excitations of the gauge bosons, as well as by the presence of bulk 
higher-dimensional operators. 
This dynamical mechanism results in a composite Higgs, which is highly localized  
and generally heavy. The localization of fermions in the five-dimensional bulk 
naturally leads to the standard model Yukawa couplings via the action of 
the bulk higher-dimensional operators, which are suppressed by the Planck scale. We 
obtain the spectrum of the model and explore some of its phenomenological consequences, 
both for electroweak precision constraints as well as at the Large Hadron Collider.   
}


\begin{document}


\section{Introduction} \setcounter{equation}{0}
\label{intro}
The standard model (SM) is an extremely successful description of the
electroweak interactions. However, the instability of the weak scale under radiative 
corrections leads us to believe that there should be physics beyond the SM at an energy 
scale not far beyond the TeV. The origin of electroweak symmetry breaking (EWSB) 
as well as of fermion masses, might be associated with this new dynamics. 
For instance, the fact that the 
top mass is of the order of the weak scale suggests that its origin 
might be associated with EWSB. This was first proposed in Ref.~\cite{bhl}, where 
the condensation of the top quark leads to its dynamical mass and the breaking of the 
electroweak symmetry. The new dynamics responsible for the condensation cannot be far above 
the weak scale in order to avoid fine-tuning. However,  
in order for the new dynamics to occur close to the TeV scale, the top-quark's dynamical mass 
would have 
to be considerably larger than what it actually is, about $(600-700)~GeV$. 
Conversely, in order to obtain the experimentally observed $m_t$, the new 
dynamics has to reside at a scale of about $10^{15}$~GeV or so. Already in Ref.~\cite{bhl} it 
was pointed out that the condensation of a fourth generation, with TeV dynamics, could be the 
solution  to this problem. In this context, however, it is not simple to arrange the dynamics 
so that only the fourth generation condenses, as well as to explain
the rest of the fermion 
masses~\cite{holdom1}. 
The strong dynamics appears unnaturally selective of the different fermion generations.

Non-trivial strong dynamics in 4D can be described by a weakly-coupled theory in 5D through  
holography~\cite{holorefs}. Of particular interest is the case in which the metric of the extra 
dimension is Anti--de-Sitter (AdS). This choice leads to the possibility of 
large separation of energy 
scales  and has been proposed to solve the hierarchy problem~\cite{rs}. 
In a compact extra dimension with AdS metric, 
and the fundamental scale being the Planck mass $M_P$, 
it is possible to generate the TeV scale at a distance $\pi R$ from the origin, as long as 
$k R \sim (10-12)$, where $k$ is the AdS curvature $k\sim M_P$.  This type of setup not only 
can explain the hierarchy between $M_P$ and the weak scale. If fermions are allowed in the 
bulk~\cite{gp,gn,cn,dhr}, 
their localization in the extra dimension, determined by their $O(M_P)$ bulk masses, results 
in exponentially separated overlaps with the TeV scale, which would explain the fermion mass hierarchy naturally. 
In this scenario the Higgs field must remain highly localized close to the TeV brane in order not to 
receive quadratic divergences to its mass above the TeV scale. In principle, this localization 
should have a dynamical origin. At the moment there is only one dynamical mechanism localizing 
the Higgs field. This naturally occurs in models where the Higgs is obtained from a higher dimensional 
component of a bulk gauge field~\cite{cnp,acp}. 
In this case, the Higgs corresponds to the zero mode of the part of the bulk gauge field related to 
the broken generators. This essentially means that this Higgs is a pseudo-Nambu--Goldstone boson. 
Alternatively, Higgsless scenarios~\cite{higgsless} have been proposed
in AdS$_5$, where the electroweak symmetry is broken by boundary
conditions. Finally, it is possible to interpolate between these two
pictures~\cite{gaugephobic} by having a bulk Higgs with a TeV-localized vacuum expectation value
(VEV). 

In this paper, we consider four SM generations propagating in an AdS$_5$ bulk. 
The bulk masses of the fourth-generation are chosen so as to localize its zero-modes towards
the TeV brane. This in turn induces strong couplings of the fourth-generation to the 
Kaluza-Klein (KK) excitations of the gauge bosons, particularly of the fourth-generation quarks 
with the KK gluons. Also, the inevitable presence of bulk higher-dimensional operators induces
additional zero-mode four-fermion operators. 
The effectively induced four-fermion interactions 
can be super-critical, breaking chiral symmetry and the electroweak symmetry. 
In this realization of the 
fourth-generation condensation, the obtained dynamical fermion mass is approximately 
$(600-700)~$GeV, for 
KK gauge masses in the few TeV region.  
In the simplest realization, with only one fourth-generation zero-mode quark  
condensing (e.g. the up-type), 
the effective theory at energies below the KK mass scale
presents a spectrum containing only one
composite scalar doublet corresponding to the Higgs field. As long as the four-fermion interactions
induced by the KK excitations are super-critical in the condensing channel, 
the Higgs acquires a VEV, 
and the electroweak  symmetry is broken, giving masses to the $W^\pm$ and $Z^0$. 
The TeV localization of the Higgs field is a direct result of the localization of its constituents.
We find typically a heavy Higgs, as is to be expected due to its
highly TeV-localized wave-function, about $900~$GeV for a few TeV KK masses.

Bulk four-fermion operators, suppressed by the Planck scale, will be responsible for 
fermion masses.
In particular, four-fermion operators involving the condensing fourth
generation quarks will result in fermion masses. Just as in any bulk Randall-Sundrum (RS)
model with a TeV-localized Higgs, fermion zero modes with large overlap with the TeV brane will 
be heavier (e.g. the fourth generation, the top quark), whereas Planck-brane localized 
fermions will have suppressed coefficients in the four-dimensional effective operators resulting
from the higher dimensional 5D operators. Thus, the model maintains the natural generation of the 
fermion mass hierarchy, a very compelling feature of bulk RS models.

The model we present here is a realization of the flavor-dependent strong dynamics 
necessary in a fourth-generation condensation scenario, in the context of the AdS/CFT correspondence. 
It also provides an alternative way to localize the Higgs field close to the TeV brane in 
RS models, other than the one proposed in Refs.~\cite{cnp,acp}. Also,
unlike in the model of Ref.~\cite{gaugephobic}, the Higgs VEV and its
localization are not free parameters, but are fixed by the dynamics of
the fourth-generation in the bulk. 
The general idea allows for various choices, from the number of condensing fermions, to the 
presence of a right-handed neutrino zero-mode, and generally the choice of fermion representations
under the bulk gauge group. We will try to be as definite and simple as possible, leaving alternatives
for further work.

The plan for the paper is as follows: in the next Section we present the model and 
show how electroweak symmetry and fermion masses arise in it.  
In Section~\ref{rge}, 
the low energy effective theory for the 
zero modes and the Higgs is built. We compute the masses of the fourth-generation 
fermions and the Higgs making use of renormalization group methods. 
In Section~\ref{ewpc} we consider the electroweak precision constraints on the model, and 
its main phenomenological features, especially at colliders, are discussed in Section~\ref{pheno}. 
We conclude in Section~\ref{conc}. 

\section{The Model}
\label{themodel}

\subsection{The Five-dimensional Setup}
\label{fdsetup}

We consider a theory with one compact extra dimension where the metric is given by~\cite{rs} 
\be
ds^2 = e^{-2\,k\,y} \eta_{\mu\nu} dx^\mu dx^\nu - dy^2~,
\label{metric}
\ee
and $k \sim M_P$ is the AdS curvature. The orbifold compactification $S_1/Z_2$ results in 
a slice of AdS in the interval $[0,\pi R]$, with $R$ the compactification radius.
In order for the weak  scale to arise at the brane in $y=\pi R$, we need $k\sim 11$. 
All fermions propagate in the 5D bulk.  These include the standard three generations, as well as
a complete fourth generation. 
The boundary conditions are such that the zero-mode spectrum reproduces that of the 
SM fermions, with the addition of one extra SM generation. 
The gauge symmetry in the bulk cannot be just the SM, since the 
custodial breaking contributions from the $U(1)_Y$ KK modes would result in unacceptably large 
isospin violation. Instead, we consider that the bulk gauge theory is~\cite{adms}
$SU(2)_L\times SU(2)_R\times U(1)_{X}$, where the boundary conditions in the UV lead to 
$SU(2)_R\times U(1)_{X}\to U(1)_Y$. Additional symmetries  may be imposed in order to 
protect the $Z\bar b b$ coupling from large corrections~\cite{acdp}. When this is the case, 
third generation fermions have to be in specific representations of the gauge group. 
For instance, left-handed quark doublet must be a $\bf(2,2)_{2/3}$ under 
$SU(2)_L\times SU(2)_R\times U(1)_X$. On the other hand the field with $t_R$ as its zero mode, 
can be in either $\bf(1,1)_{2/3}$ or ${\bf(3,1)_{2/3}}\oplus{\bf(1,3)_{2/3}}$. 
We will take similar representations for the fourth
generation. Regarding leptons, we
will assume at a minimum the presence of a fourth generation lepton
bi-doublet, and a singlet with a charged right-handed zero-mode
$E_R$. Also, if we assume that the fourth-generation neutrino has a
large Dirac mass, there should be an additional  bulk field with a
right-handed neutrino zero-mode $N_R$. 
Since in this paper we are mainly concerned with the fourth-generation zero
modes, we will not need to make a choice of bulk representation,
whenever such choice is possible.

Bulk fermion masses are naturally of the order of the AdS curvature $k$, such that 
\be
M_f = c_f\,k~,
\ee
with $c_f\sim O(1)$. They determine the localization of fermion zero modes in the 
bulk.
The localization of the fourth-generation zero-modes very close to the 
TeV brane
results in strong interactions with the gauge boson KK modes. 
The couplings of fermions to KK gauge bosons are generically determined
from the expression for the 5D coupling
\be
g_5\,\int d^4x\, \int^{\pi R}_0 dy \sqrt{g} \bar\Psi(x,y)\,e^{ky}\,\gamma^\mu T^a 
\Psi(x,y) A^a_\mu(x,y)~.  
\label{fdcoupling}
\ee
where the factor of $e^{ky}$ comes from the vierbein, 
$g_5$ is the 5D gauge coupling, and the $T^a$ are the generators of the 
gauge symmetry. 
Expanding $A_\mu(x,y)$ and $\Psi(x,y)$ in their KK modes as
\be
A_\mu(x,y) = \frac{1}{\sqrt{\pi R}}\, \sum_n\, \chi_n(y) A^{(n)}_{\mu}(x) ~, 
\ee
and 
\be
\Psi(x,y) = \frac{1}{\sqrt{\pi R}}\, \sum_n\, e^{2ky}\,f_n(y) \psi^{(n)}(x) ~, 
\ee
and integrating over the compact dimension
we obtain the coupling of the $i$th fermion KK mode to the $n$-th KK mode of the gauge boson:
\be
g_{in} = \frac{g}{\pi R}\, \int_0^{\pi R} dy \,e^{ky}\,\left| f_i(y)\right|^2 \, \chi_n(y) ~, 
\label{gin} 
\ee
where we have defined the 4D gauge coupling by $g = g_5\,/\sqrt{\pi R}$. 
Here, we are interested in the couplings of the fermion zero-modes with the 
first KK excitations of the gauge bosons. 
The wave functions for the nth KK gauge boson is given by~\cite{gp,gn}:
\be
\chi_n(y) = \frac{e^{ky}}{N_n}\,\left[ J_1(\frac{m_n}{k}\,e^{ky}) + \alpha_n 
Y_1(\frac{m_n}{k}\,e^{ky})\right]~,
\label{kkgauge}
\ee
where $m_n$ is the mass of the nth KK excitation of the gauge boson, $N_n$ is the normalization,  
$J_1$ and $Y_1$ are Bessel functions, and the constant $\alpha_n$ is defined 
by 
$\alpha_n = -J_0(\frac{m_n}{k})/Y_0(\frac{m_n}{k})= 
-J_0(\frac{m_n\,e^{k\pi R}}{k})/Y_0(\frac{m_n\,e^{k\pi R}}{k})$, which also determines the 
KK masses $m_n$.

For the zero mode fermions, one obtains the wave-functions
\be
f_0^{L,R}(y) = \sqrt{\frac{k\pi R(1\mp 2 c_{L,R})}{e^{(1\mp 2 c_{L,R}) k\pi R} - 1}}\,
e^{\mp c_{L,R} \,ky}~.
\ee
However, it is more useful to consider 
the $y$-dependence of the kinetic terms of the KK fermions as the
effective fermion wave-functions. In this case the $y$ dependence is
\be
\hat{f}_{L,R}(y) = \frac{e^{(\frac{1}{2}\,\mp \,c_{L,R})ky}}{N_{L,R}}~,
\label{wflh}
\ee
where we defined the normalization factors
\be
\frac{1}{N_{L,R}} \equiv \sqrt{\frac{1\mp 2 c_{L,R}}{e^{k\pi R(1\mp 2 c_{L,R})}-1}} 
\ee
Then, left-handed (right-handed) fermions with $c_L>1/2$ ($c_R< -1/2$) are localized towards
the Planck brane, whereas left-handed (right-handed) fermions with  $c_L<1/2$ ($c_R> -1/2$)
are localized close to the TeV brane. 
Light fermions are of the first kind, while heavier fermions must be localized near the TeV brane. 
This is the case with the top quark, and now in this model also with all the fourth-generation
fermions. 

For values of the bulk mass parameter $c_L>1/2$, the zero-mode fermion 
couples universally, as well as weakly, to the first KK gauge boson. For $c_L<1/2$, 
the coupling can be considerably enhanced above the gauge coupling. We will consider
that the fourth generation has bulk mass parameters that localize it very close to the 
TeV brane, and therefore it will have very strong couplings to the KK gauge bosons.

In addition to the strong interactions among fourth-generation zero-modes induced 
by the KK gauge bosons, there are interactions induced by bulk higher-dimensional 
operators. Of particular interest are  the four-fermion bulk operators
\be
\int dy\,\sqrt{g}\,\frac{C^{ijk\ell}}{M_P^3}\,\bar{\Psi}^i_L(x,y) \Psi^j_R(x,y) 
\bar{\Psi}^k_R(x,y) \Psi_L^\ell(x,y)
\label{ffermion5d}~, 
\ee
where $C^{ijk\ell}$ are generic coefficients, with $i,j,k,\ell$ standing for  generation 
indices as well as other indices such as isospin, 
and the $\Psi(x,y)$'s can be bulk quarks or leptons.
The naive dimensional analysis (NDA) estimate of the 5D coefficients in (\ref{ffermion5d}) 
gives
\be
C^{ijk\ell} \sim \frac{36\,\pi^3}{N}~,
\label{nda} 
\ee
with $N$ the number of fermion flavors that can be accommodated in a loop. 
If we assume that all the $C$'s are of the same order, then $N\sim O(100)$\footnote{
For instance, if all the $C's$ are exactly equal, the number of fermions inside a loop mediating any given 
four-fermion interaction is $N=80$.}. 
These bulk operators lead to 4D four-fermion 
operators involving the various fermion zero and KK modes. The four-fermion 
interactions induced among fermion zero-modes are given by~\cite{gp} 
\be
C^{ijk\ell}\,\frac{k}{M_P^3}\, 
\, \frac{e^{k\pi R(4 - c_L^i - \tilde{c}_R^j - \tilde{c}_R^k - c_L^\ell)} - 1}
{4 - c_L^i - \tilde{c}_R^j - \tilde{c}_R^k - c_L^\ell}\,\,
\frac{\bar{\psi}_L^{i(0)} \psi_R^{j(0)} \bar{\psi}_R^{k(0)}
\psi_L^{\ell(0)}}
{N^i_L\,N^j_R\,N^k_R\,N^\ell_L}~,
\label{ffopin4d}
\ee
where we defined $\tilde{c}_R^i = -c^i_R$ for convenience. 

Once again, localization is determining the size of these contributions. For instance, for 
bulk mass parameters for two of the fermions satisfying $(c_L^i,\tilde{c}_R^j) > 1/2$, 
i.e. two of the fermions being Planck localized, these four-fermion operators are exponentially 
suppressed. On the other hand if all four-fermions have bulk mass
parameters localizing the 
zero-modes 
towards the TeV brane ($(c_L,\tilde{c}_R) < 1/2$), the corresponding contribution will be 
only suppressed 
by the TeV scale. In particular, the four-fermion interactions induced among 
fourth-generation zero-modes are only TeV suppressed, with a dimensionless coupling of the form
\be
\sim C^{4444}\,\left(\frac{k}{M_P}\right)^3\,
\frac{(1-2c_L^4)(1-2\tilde{c}_R^4)}{2\,(2 - \tilde{c}_R^4 - c_L^4)}~.
\label{fgenc}
\ee 
were the sub-indices in the coefficients denote flavor quantum numbers.
\begin{figure}
\begin{center}
\epsfig{file=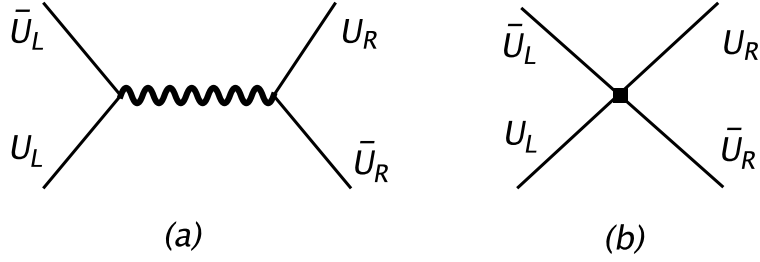,width=10cm}
\caption{Two contributions to four-fermion interactions of the up-type fourth-generation quark: 
(a) from the interactions with a KK gluon;  
(b) from the four-fermion interactions induced by 
the bulk operators of (2.11).}
\end{center}
\label{fig:fourfermion}
\end{figure}

\subsection{Four-fermion interactions and Electroweak Symmetry Breaking}
\label{sub:ffermion}

We now examine the four-fermion interactions among fermions
induced by the exchange of KK gauge bosons. The  strongest coupling of the fourth 
generation is that of the first KK gluon to the fourth generation quarks.  
For instance, considering the zero-mode $U$ quark we have the following 
four-fermion interaction below the mass of the first excitation of the KK gluon, $M_{KK}$:
\be
- \frac{g_{01}^L\,g_{01}^R}{M_{KK}^2}\,\left(\bar{U}_L \gamma_\mu T^A U_L\right)\, 
\left(\bar{U}_R \gamma_\mu T^A U_R\right)~, 
\ee 
where $U$ is the zero mode of the fourth-generation up-type quark, 
$g_{01}^L$ and $g_{01}^R$ are the left-handed and right-handed $U$  
couplings to the first KK gluon excitation, and $T^A$ are the usual $SU(3)_c$ generators.
After Fierz rearrangement, we can re-write this interaction as 
\be
\frac{g_{01}^L\,g_{01}^R}{M_{KK}^2}\,\left\{ \bar{U}^a_L U^a_R \,\,\bar{U}^b_R U^b_L 
- \frac{1}{N_c}\, \bar{U}^a_L U^b_R \,\,\bar{U}^b_R U^a_L \right\}~, 
\label{ffer2terms}
\ee
where $a,b$ are $SU(3)_c$ indices.
The color singlet term in (\ref{ffer2terms}) is attractive, whereas the color octet is 
repulsive, as well as suppressed by $1/N_c$. We then concentrate in the color singlet 
four-fermion interaction. 
Likewise, electroweak KK gauge bosons give similar, although much smaller, contributions. 
We will neglect them in what follows. However, we must take into account the contributions generated
from bulk higher-dimensional operators such as (\ref{ffermion5d}), since they 
are generally comparable  to the ones obtained from KK exchange. Both these contributions, depicted 
in Figure~\ref{fig:fourfermion}, result in an effective four-fermion interaction of the fourth-generation 
quarks. For the $U$ quark, for instance, we then have an effective four-fermion coupling
\be
g_{U}^2 \equiv g_{01}^L\,g_{01}^R + x_1^2\,C^{4444}_{uu}\,\left(\frac{k}{M_P}\right)^3\,
\frac{(1-2c_L^4)(1-2\tilde{c}_R^4)}{2(2-c^4_L-\tilde{c}^4_R)}~,
\label{gueff}
\ee
where $x_1 \equiv M_{KK}/\Lambda_{\rm TeV}$, is the mass of the first KK gauge bosons in units of the 
TeV scale defined as $\Lambda_{\rm TeV} = k\,e^{-k\pi R}\sim O(1)$~TeV.

There is a value of $g_U^2$ above which a condensate forms
\be
\langle \bar{U}_L\,U_R\rangle \neq 0~,
\ee
leading to electroweak symmetry breaking and dynamical masses for the 
condensing fermions. It is possible to obtain the criticality condition on the coupling
from a gap equation. Here we want to describe electroweak symmetry breaking 
through the vacuum expectation value of the ensuing composite scalar, the Higgs. 
For this purpose, we start from the Lagrangian
\be
{\cal L}  = \bar{U}_L i\not \hspace{-0.1cm}D U_L + \bar{U}_R i\not \hspace{-0.1cm}D U_R
+ \frac{g_{U}^2}{M_{KK}^2}\,\left( \bar{U}_L U_R \,\,\bar{U}_R U_L \right) ~,
\label{njlfermions}
\ee
This can be re-written as 
\be
{\cal L} = \bar{U}_L i\not \hspace{-0.1cm}D U_L + \bar{U}_R i\not \hspace{-0.1cm}D U_R 
+ g_U\,\bar{Q}_L H U_R - M_{KK}^2 H^\dagger H + {\rm h.c.} ~,
\ee
where $Q_L^T\equiv (U_L~ D_L)^T$, $H$ is a non-propagating $SU(2)_L$ doublet, and we have omitted the down-type 
quark kinetic terms. At scales $\mu < M_{KK}$, $H$ develops a kinetic term as well as 
a self-coupling, resulting in 
\bear
{\cal L}(\mu) &=& Z_{U_L}\,\bar{U}_L i\not \hspace{-0.1cm}D U_L 
+ Z_{U_R}\,\bar{U}_R i\not \hspace{-0.1cm}D U_R + \cdots   
+ Z_{g_U} g_U\,\bar{Q}_L H U_R +h.c. \nonumber\\ 
&& +Z_H (D_\mu H)^\dagger D^\mu H
- m_H^2 H^\dagger H -\frac{\lambda}{2}\,\left(H^\dagger H\right)^2~,
\label{mulag}
\eear
where the wave-function renormalizations $Z_{U_L}$, $Z_{U_R}$, $Z_H$,$\dots$, as well as 
$Z_{g_U}$, $m_H$ and $\lambda$, can be easily computed in the one loop approximation.  
For instance the dominant contribution to $m_H$ results in 
\be
m_H^2 = M_{KK}^2\left( 1 - \frac{g_U^2\,N_c}{8\pi^2}\right) +\cdots~.
\label{mH2}
\ee
Thus, we see that the effective potential for $H$ at low energies develops a non-trivial
vacuum if 
\be
g_U^2 > \frac{8\pi^2}{N_c}~.
\label{criticality}
\ee
This condition is easily satisfied in these models, even in the absence of 
the four-fermion operators of eqn.~(\ref{ffermion5d}), by giving enough 
localization to the fourth-generation. For instance, if $(c_L^4,\tilde{c}_R^4)<0$  
the KK-gluon induced four-fermion interactions are always super-critical.
In addition, if we include the effects of KK fermions in the effective Higgs theory, 
the resulting critical coupling would be lower than the one obtained in (\ref{criticality}).
In any case, the exact value of the critical coupling is not important for the calculation
of the spectrum in this model.

Equation~(\ref{criticality}) coincides with the criticality condition obtained by making a one-loop 
gap equation analysis of (\ref{njlfermions}). Then, we see that if the couplings 
of zero-mode fermions to KK gauge bosons are strong enough, they could lead to 
electroweak symmetry breaking. Among the SM fermions, the best candidate for accomplishing 
this is the top quark, as in top-condensation models~\cite{bhl,tc1,tc2}. 
However, even if we assumed that the effective four-fermion interactions of top quarks
were super-critical, this would lead either to a top mass that is too large, or to 
a cutoff that has to be above $10^{15}$~GeV.
Earlier attempts to embed top-condensation in flat~\cite{bogdan} and 
warped~\cite{nuria} extra-dimensional theories, required the condensation of a large number
of KK fermions in order to obtain the correct value of $m_t$. In the present AdS$_5$ 
setup this is very difficult to achieve and requires unnaturally large values for the 
coefficients in (\ref{ffermion5d}), given that higher KK modes have weaker couplings.
A fourth generation with zero modes highly localized towards the TeV brane is guaranteed to condense. 
For simplicity, we will consider here the case where only the up quark $U$ condenses. The 
case with the $D$ also condensing leads to a more complicated scalar sector~\cite{fgcon}.   

The coefficient of the kinetic term of $H$, and its self-coupling, computed at one loop, are
given by
\bear
Z_H &=& \frac{g_U^2 N_c}{16\pi^2}\,\ln\left(\frac{M_{KK}^2}{\mu^2}\right)~,\label{zh}\\
\lambda &=& \frac{g_U^4\,N_c}{8\pi^2}\,\ln\left(\frac{M_{KK}^2}{\mu^2}\right)~,\label{lambda}
\eear
where we have only included the up quark zero mode contributions. We notice that 
both $Z_H$ and $\lambda$ vanish at the cutoff $\Lambda=M_{KK}$, reflecting the 
compositeness conditions.  
Completing the renormalization procedure, we consider the scalar contributions to $Z_{U_L}$ 
and $Z_{U_R}$, as well as the coupling renormalization $Z_{g_U}$ coming from scalar exchange.  
After the replacements
\be
Z^{1/2}_{U_L} U_L \to U_L~,    ~~~~~~~  Z^{1/2}_{U_R} U_R \to U_R~, ~~~~~~~Z^{1/2}_H H \to H
\label{renfields}~, 
\ee
and the definition of the renormalized quantities
\be
\bar{m}_H^2 = \frac{m_H^2}{Z_H}, ~~~~~~~~~
\bar{\lambda} = \frac{\lambda}{Z_H^2}, ~~~~~~~~~
\bar{g}_U = \frac{Z_{g_U}}{\sqrt{Z_{U_L}\,Z_{U_R}\,Z_H}}\,g_U~,
\label{rencouplings}
\ee
the renormalized lagrangian reads
\bear
{\cal L}_r &=& \bar{U}_L i\not \hspace{-0.1cm}D U_L 
+ \bar{U}_R i\not \hspace{-0.1cm}D U_R + \cdots   
+ \bar{g}_U\,\bar{Q}_L H U_R +h.c. \nonumber\\ 
&& + (D_\mu H)^\dagger D^\mu H
- \bar{m}_H^2 H^\dagger H -\frac{\bar{\lambda}}{2}\,\left(H^\dagger H\right)^2~.
\label{renlag}
\eear
Assuming that the criticality condition (\ref{criticality}) is satisfied, the Higgs 
field $H$ acquires a VEV
\be
\langle H \rangle = \left(\begin{array}{c} 
 v/\sqrt{2}\\ 0 \end{array} \right)~,
\label{hvev} 
\ee
giving the condensing fermion a dynamical mass $m_U = \bar{g}_Uv/\sqrt{2}$. 
Here we take $v\simeq 246$~GeV, which results in the correct value for
$M_W$ at this order in perturbation theory. 
The Higgs mass, $m_h = \sqrt{\bar{\lambda}} v$, can be computed in this approximation and 
satisfies the Nambu--Jona-Lasinio (NJL) relation, $m_h = 2\,m_U$. However, this simplistic 
prediction receives important corrections that will be addressed in the next section. 
The same can be said of the prediction for the dynamical fermion mass $m_U$, which at this 
level of accuracy must satisfy the Pagels--Stokar expression
\be
v^2 = m_U^2\,\frac{N_c}{8\pi^2}\,\ln\left(\frac{M_{KK}^2}{m_U^2}\right)~,
\label{psformula}
\ee
which points to dynamical masses in the few hundred GeV for the up-type fourth generation 
quark. We will refine the predictions for the dynamical fermion masses and the Higgs mass
in Section~\ref{rge}, where we will make use of the full renormalization group 
evolution of the couplings $\bar{g}_U$ and $\bar{\lambda}$. 
But before that, we will address the origin of the masses of all other (non-condensing) fermions.  

\subsection{Fermion Masses}
\label{fermionmasses}

In the previous section we have shown that if the zero modes of quarks of the fourth generation
are localized enough, they can condense and break the electroweak symmetry. 
The condensate $\langle \bar U_L U_R\rangle$ results in a dynamical mass for the $U$ quark
zero mode. 
Higher-dimensional operators suppressed by $M_P$ such as (\ref{ffermion5d}), will 
result in masses for all other zero-mode fermions upon condensation of the fourth-generation up-type 
zero-mode quark $U$, as well as of any other condensing fermion.
For instance, we could imagine that flavor was a gauge symmetry
and it was broken at the Planck scale. The same might be true of 
other symmetries, which may couple quarks and leptons at the very high scale. 
 
When two of the zero-mode fermions are condensing, for instance $U_L$, and $U_R$ in the model we are 
considering here, the operator in (\ref{ffermion5d}) results in masses
for the other  
zero modes. This corresponds to $k=\ell=4$. 
Light fermion masses result from fermions with bulk mass parameters $(c_L^i,\tilde{c}_R^j) > 1/2$.  
Assuming  the condensate satisfies $\langle \bar U_R U_L\rangle \sim m_U^3$, these are
\bear
m_{ij} &=& C^{ij44}\,\left(\frac{k}{M_P}\right)^3\,\left(\frac{m_U}{\Lambda_{\rm TeV}}\right)^2\, 
\frac{\sqrt{(2c_L^i-1)(2\tilde{c}_R^j-1)}\,\sqrt{(1-2c_L^4)(1-2\tilde{c}_R^4)}}
{4 - c_L^i - \tilde{c}_R^j - \tilde{c}_R^4 - c_L^4}\,\nonumber\\
&&~~~~~~~~~~\times e^{k\pi R(1-c_L^i-\tilde{c}_R^j)}\, m_U~.
\label{lightmass}
\eear
 The masses in 
(\ref{lightmass}) are exponentially suppressed and lead to light fermion masses. 
On the other hand, for $(c_L^i,\tilde{c}_R^j) < 1/2$, we arrive at 
\be
m_{ij} = C^{ij44}\,\left(\frac{k}{M_P}\right)^3\,\left(\frac{m_U}{\Lambda_{\rm TeV}}\right)^2\, 
\frac{\sqrt{(1-2c_L^i)(1-2\tilde{c}_R^j)}\,\sqrt{(1-2c_L^4)(1-2\tilde{c}_R^4)}}
{4 - c_L^i - \tilde{c}_R^j - \tilde{c}_R^4 - c_L^4}\,m_U~, 
\label{heavymass}
\ee
which is un-suppressed and of order $m_U$, up to factors of $O(1)$. This is the case for the top
quark, the fourth-generation quark $D$, as well as the fourth-generation 
leptons. Thus, all fourth-generation fermions have masses in the several hundred GeV, with 
their exact values depending on the details of their localization near the TeV brane. 
This picture of fermion masses is consistent with the one obtained with a TeV-brane-localized 
Higgs~\cite{gp,adms}. Then, in the effective theory with a composite Higgs field described in the 
previous section, it is possible to obtain the observed 4D Yukawa couplings starting from these 
four-fermion interactions. \\

A more precise prediction for the fourth-generation masses, as well as for the mass 
of the Higgs boson, can be obtained by considering the full renormalization group running. 
We do this in the next section.

\section{Renormalization Group Effects and Mass Predictions}
\label{rge}

In order to obtain better predictions for the spectrum of the theory at low energies, 
we must consider the effects of the renormalization group running, especially on the 
Yukawa coupling of the fourth-generation up quark $U$, and the (renormalized) 
Higgs self-coupling $\bar\lambda$. Here we follow closely Bardeen, Hill and Lindner~\cite{bhl,hs}. 

\subsection{Yukawa Running and Dynamical Fermion Mass}
\label{yukawarge}

\begin{figure}
\begin{center}
\psfrag{m}[b]{$m_U,~m_h$~[TeV]}
\psfrag{L}[t]{\hspace*{1cm}$\Lambda$ ~[TeV]}
\epsfig{file=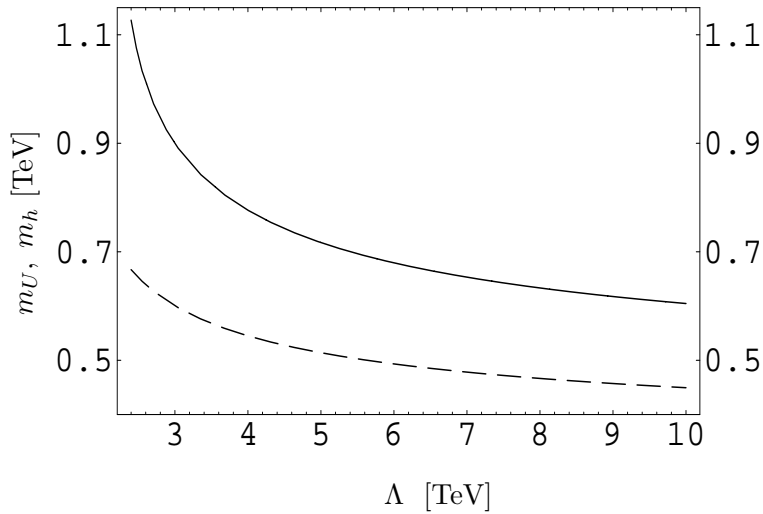,width=10cm,height=7cm,angle=0}
\caption{Dynamical mass of the fourth-generation up quark, $m_U$
(dashed-line); and
the physical Higgs mass $m_h$ (solid line), both vs. the cutoff $\Lambda$, in TeV.}
\label{massvslam} 
\end{center}
\end{figure}

Considering the one-loop contributions of the Yukawa coupling $\bar{g}_U$ to the wave-function 
renormalizations $Z_{U_L}$ and $Z_{U_R}$, as well as the contributions to $Z_{g_U}$ and $Z_H$, 
one obtains, neglecting gauge interactions, 
\be
\frac{d\bar{g}_U}{dt} = \frac{\bar{g}_U^3}{16\pi^2}\left[ \frac{3}{2} + N_c\right]~,
\label{yrgenogauge}
\ee 
where $t=\ln(\mu)$ and $\mu$ is the renormalization scale. At high energies, the Yukawa 
coupling blows up. At low energies, however, the gauge contributions 
are important. If we take them into account we have 
\be
\frac{d\bar{g}_U}{dt} = \frac{1}{16\pi^2}\left[ \frac{9}{2}\,\bar{g}_U^3 - C(t) \,\bar{g}_U]\right]
~,\label{yrenwgauge}
\ee
where the $SU(3)_c\times SU(2)_L\times U(1)_Y$ running couplings are taken into 
account in 
\be
C(t) = 8g_s^2(t) + \frac{9}{4} g^2(t) + \frac{17}{12} g'^2(t)~,
\label{runinigcs}
\ee
and we will use the values of the couplings as extracted in the particle data book~\cite{pdg}. 
Equation~(\ref{yrenwgauge}) is solved with the boundary condition 
\be
\bar{g}_U \to \infty, ~~~~~~~~~~~{\rm for} ~~~\mu\to\Lambda~,
\label{bc}
\ee
where $\Lambda$ is the cutoff, and we take $\Lambda=M_{KK}$. The main effect is from the QCD 
coupling $g_s(t)$. The solution for the physical mass $m_U$ comes from 
$m_U = \bar{g}_U(m_U)\,v/\sqrt{2}$. In Figure~\ref{massvslam} we show the result for the dynamical 
fourth-generation mass $m_U$ as a function of the cutoff $\Lambda$. 
Since the cutoff is $\Lambda = M_{KK} = O({\rm few})$~TeV, we predict the dynamical fourth-generation 
mass in the range $m_U\sim (600-700)~$GeV, somewhat lower than the naive prediction of the
previous section in (\ref{psformula}).  Potentially large mixing of fourth-generation zero-mode fermions with 
their KK modes might lower their masses even further, perhaps as much as $30\%$~\cite{adsewpt}. 

The masses of the other fourth-generation fermions depend on the choices of their bulk mass
parameters, but they  are typically of the order of $m_U$. Thus, we will be able to 
choose the amount and sign of isospin violation introduced by the zero-mode fourth generation. 

\subsection{The Higgs Mass}
\label{hmass}

The renormalized Higgs self-coupling determines the Higgs mass through 
$m_h = \sqrt{\bar{\lambda}(m_h)}\,v$. If we neglect the gauge interactions and only consider
the effect of the fourth and third generation Yukawa couplings, 
the renormalization group evolution of $\bar\lambda$ is given by 
\be
\frac{d\bar\lambda}{dt} = \frac{12}{16\pi^2}\left[ \bar{\lambda}^2 + \sum_{f}\bar{g}_f^2 \bar\lambda 
-\sum_{f}\bar{g}_f^4\right]~,
\label{lamnogauge}
\ee 
where $\bar{g}_f$ is the Yukawa coupling of the fermion $f$, the sums are over the fourth-generation quarks 
and a lepton doublet, and also include the top quark.  
From the expressions (\ref{lambda}) and (\ref{rencouplings}) for $\bar\lambda$, one would only 
obtain the last two terms in (\ref{lamnogauge}).  However, from these expressions we also see 
that $\bar\lambda$ diverges at the cutoff $\Lambda = M_{KK}$. It is then consistent to 
consider the presence of the tree-level $\bar\lambda$ interaction that gives rise to the first term 
in (\ref{lamnogauge}). This is then the same RGE as for the SM Higgs self-coupling. 

The solutions of (\ref{lamnogauge}) must satisfy compositeness conditions, as determined by eqns.~(\ref{zh}) and 
(\ref{lambda}). We may cast these 
by making  the replacement $H\to \bar{g}_U\,H$ in (\ref{renlag}), the renormalized lagrangian. 
This results in an effective Higgs self-coupling that goes like
$\bar\lambda/\bar{g}_U^4$,   
which should go to zero at the cutoff $\Lambda$, to satisfy compositeness. 
Then, we see that $\bar\lambda$ must diverge slower than  $\bar{g}_U^2$. 
This implies that the solutions to (\ref{lamnogauge}) flow to an 
ultra-violet fixed-point~\cite{bhl}, such that
\be
\bar\lambda \simeq \bar{g}_U^2\,x_+~,
\label{lamsolution}
\ee
with $x_+= (\sqrt{65}-1)/8\sim 0.88$, and where we have considered
that all the fourth-generation Yukawas are of order $\bar{g}_U$.  Making use of our results for 
$\bar{g}_U$, this results in 
\be
m_h \simeq 1~{\rm TeV}~,
\label{mhnogauge}
\ee
for a cutoff $\Lambda\sim 2.5~$TeV, which is still very close to the naive NJL prediction. 

Considering the electroweak gauge corrections, the full RGE for 
$\bar\lambda$ now is 
\be
\frac{d\bar\lambda}{dt} = \frac{12}{16\pi^2}\left[ \bar{\lambda}^2 + 
\left(\sum_{f}\bar{g}_f^2 -A(t)\right)\bar\lambda 
+B(t)-\sum_f\,\bar{g}_f^4\right]~,
\label{lamwgauge}
\ee
where 
\bear
A(t) &=& \frac{1}{4} g'^2(t) + \frac{3}{4}g^2(t)\nonumber\\
B(t) &=& \frac{1}{16} g'^4(t) +\frac{1}{8} (g(t) g'(t))^2 + \frac{3}{16}g^4(t)~. 
\eear
As we can see from the Figure~\ref{massvslam}, the addition of the gauge contributions
does not modify the prediction for $m_h$ greatly. This remains a very heavy Higgs, if the 
cutoff is kept not far above the TeV scale.

\section{Electroweak Precision Constraints}
\label{ewpc}
Bulk RS models on which we based our construction, have an enlarged isospin symmetry 
given by the extension from the SM gauge group to $SU(2)_L\times SU(2)_R\times U(1)_{X}$. 
This forbids tree-level contributions to the $T$ parameter. On the other hand, 
there are important contributions to the 
$S$ parameter already at tree-level~\cite{adms}.  
These can be seen as coming from the interactions of light (Planck-localized) 
fermions with the gauge bosons, through 
the KK modes. The modified couplings, being universal, can be re-absorbed into a redefinition of the gauge 
fields, resulting in contributions to the oblique parameters $S$ and $T$. 
Particularly dangerous is the 
$S$ parameter contribution, given by 
\be
S_{\rm tree} \simeq 12\pi\,\frac{v^2}{M_{KK}^2}~.
\label{stree}
\ee
Additional tree-level contributions correspond to operators of dimension eight or higher, and 
are further suppressed by factors of $v^2/M_{KK}^2$. 

In the present model, the presence of a fourth generation induces new loop contributions, 
both from the fourth-generation zero modes, as well as their KK excitations. 
The presence of a degenerate SM fourth generation (the zero-modes) results in a positive shift of the 
$S$ parameter given by 
\be
S_{4g}\simeq \frac{2}{3\pi}~, 
\label{s4g}
\ee
with this results somewhat smaller if the down sector is lighter than the up.
Recent re-examination of the constraints on a fourth generation coming from 
electroweak precision measurements has shown~\cite{kribs} that the presence of these states is not 
in serious contradiction  with data, as it is concluded in Ref.~\cite{pdg}. 
This is particularly the case if the fourth-generation quarks have splittings giving a 
positive contribution to $T$. In our model, this can be naturally achieved by having 
the up-type quark more localized than the down-type such that $m_U > m_D$. In realizations where only the 
up quark condenses this is most easily achieved, but it can be also 
the case even if both the $U$ and the $D$ condense. 

In addition to the  SM-type contributions of the fourth-generation to $S$ and $T$, there will 
be new effects associated with the strong coupling of these states to the KK vectors. 
These effects are currently under study and will be presented elsewhere. 

The contributions of the KK fermions to $S$ and $T$ can be summed up. Their calculation is
cumbersome and we will leave it for a future publication, where we will put together all the 
electroweak precision constraints of the model~\cite{adsewpt}. However, we can already 
conclude that the main contribution to the electroweak bounds, is the tree-level contribution 
already present in Ref.~\cite{adms}, and that the presence of the bulk fourth-generation 
does not make matters much worse regarding this issue.

Also, the fact that the Higgs is heavy results in a positive shift of the $S$ parameter. 
The standard one loop contribution to $S$ from a heavy Higgs, 
results in 
\be
\Delta S_{SM}^h \simeq \frac{1}{12\pi} \ln\left(\frac{m_h}{m_h^{\rm ref.}}\right)^2~,
\label{h2ssm}
\ee
Thus, taking the reference value to be $m_h^{\rm ref.} = 114~$GeV, results in a $\Delta S\simeq + 0.1$
for the typical Higgs mass in our model.  
However, and just as for the fourth-generation contributions, the Higgs is strongly coupled and 
we must carefully study the effects of the strongly coupled KK sector on the Higgs contribution 
to $S$.

\section{Phenomenology}
\label{pheno}
The class of models presented here has a very rich phenomenology at the LHC. 
Some of its aspects will depend on the specific realization of the 
fourth-generation condensation model. For instance, the scalar sector could be richer if 
the fourth-generation down quark condenses, leading to a two-Higgs doublet spectrum. 
Also, the choice of fermion assignment to the $SU(2)_R$ group results in at least two possibilities
for the spectrum of relatively light KK fermions. 
However, there are some generic features that would constitute signals for these class of models.
If the zero-mode spectrum constitutes a complete fourth generation, its discovery at the 
LHC, in association with a heavy Higgs, would give a hint that the fourth-generation
could be associated with electroweak symmetry breaking. 
More definite proof of this, would be the observation of the strong coupling
of the fourth-generation quarks to the KK gluon excitations. 
In what follows we briefly discuss 
some generic phenomenological features of the model discussed in the previous sections. 
 
The production cross section of fourth-generation quark pairs is of about~\cite{atlastdr}  
\bear
\sigma_{Q_4\bar Q_4}&\simeq & 1~pb, {\rm ~~for} ~~~m_{Q_4} = 600~{\rm GeV}~,\nonumber\\
\sigma_{Q_4\bar Q_4}&\simeq & 0.1~pb, {\rm ~~for} ~~~m_{Q_4} = 900~{\rm GeV}~.\nonumber
\eear 
Thus, approximately $1000$ events per quark type will be produced in a typical low luminosity 
year for a $900$~GeV fourth-generation quark. However, the reach could be limited 
to masses below this due to backgrounds. 

If $m_U > m_D$, as we considered here in order to have only the $U$ quark condense, 
then for $(m_U-m_D)> M_W$, the up-type quark would decay as $U\to D W$. 
The down-type quark would decay almost exclusively to the top quark through $D\to t W$. 
Thus, the pair production of $U$ pairs results in the decay chain 
$U\bar U \to W^+W^-W^+W^-W^+W^- b\bar b$, with six $W$'s plus two $b$ jets. 
This signal has not been studied at the LHC and it appears challenging due to the large number 
of jets.  However, it appears that it might be possible to device a way to reconstruct the 
$D$ quarks, since we could use the leptonic decay of a $W$ from a $U$ decay for triggering.   
On the other hand, the $D$ pair production results in the chain $D\bar D \to W^+W^-W^+W^-b\bar b$, 
which has been studied in Ref.~\cite{atlastdr}. 
If $(m_U-m_D)<M_W$, then $U$ would decay  through $U\to b W$. Then, $U\bar U$ production is identical
to top pair production with the exception of the quark mass. A preliminary study 
in Ref.~\cite{atlastdr} shows that with $100~fb^{-1}$ luminosity it is possible to 
have a significant signal above background for masses up to at least
$700~$GeV. Other decay modes, involving significant mixing with the
third generation quarks, are studied in Ref.~\cite{holdom2}.

Regarding leptons,
the standard production cross section for a pair of charged leptons $L\bar L$ or 
of massive neutrinos $N\bar N$, is much smaller than in the quark case, since these are electroweak 
processes. Typically, for $m_L,m_N \simeq 700$~GeV, cross sections are a few $fb$. 
For instance, if $(m_L-m_N)> M_W$, the charged lepton could decay through $L\to N_L W$.  
The left-handed neutrino would then decay trough mixing with the lighter generations, through
$N_L \to \ell W$, with $\ell=\tau,\mu, e$. If these mixings are small enough, the decay might occur
outside the detector, leading to a large missing $E_T$ signal. 
If, on the other hand, $(m_L-m_N) < M_W$, then the charged lepton also must decay through 
mixing with lighter leptons, as in $L\to \nu W$. Once again, if the intergenerational mixings of the 
fourth-generation leptons are small, this could result in a slow charged track 
in the detector, which can be easily identified and might even allow the measurement of the charged
lepton mass~\cite{atlastdr}. 
Finally, if we assume the existence of a right-handed neutrino zero-mode $N_R$, for instance in order
to obtain a Dirac mass for the fourth-generation neutrino zero-mode, then its production and decay 
will depend on the transformation properties of the bulk field it belongs to. For instance, if this transforms
as a ${\bf (1,3)_0}$, it only couples to the KK excitation of the $Z'$, the state orthogonal to the SM $Z$. 
Thus, its production cross section is rather small. 
Its decays are also suppressed since they can only proceed through a 
three body decay further suppressed by the probably small mixings with the lighter lepton generations.
These events could have a very characteristic signal of large missing $E_T$ and little activity in 
the central region, albeit very rare. On the other hand, if the bulk field resulting in a fourth-generation 
right-handed neutrino transforms as a ${\bf (1,1)_0}$, the only couplings of the zero-mode are through 
the four-fermion operators of (\ref{ffermion5d}) and the effective Yukawa coupling they generate. 
Then, the operators responsible for their production and decay are effectively Planck-suppressed, making 
them possibly stable in cosmological time scales.

But the most distinct signal of the model will not be the presence of a heavy fourth generation 
in combination with a heavy Higgs. In order to clearly detect this class of models, one must 
prove that the fourth-generation is strongly coupled to the TeV scale
resonances, the gauge KK modes 
responsible for the condensation of the fourth-generation quarks. 
The main signal for this is the production of fourth-generation quarks and leptons through s-channel
production of the KK gauge bosons. 
In particular, the produced KK gauge bosons, if strongly coupled to the 
fourth-generation, would decay to it preferentially. Then, for the KK gluon for instance, we 
have that 
\be
\frac{{\rm Br}(G^{(1)} \to U\bar U)}{{\rm Br}(G^{(1)} \to t\bar t)} \sim (5-10)~,
\label{ratio42top}
\ee
depending of the parameters of the model. A careful study of the possibility of reconstructing 
these very high-invariant mass events must be done in order to evaluate how well can this 
signal be seen at the LHC. On the other hand, the contact 
four-fermion interactions coming from (\ref{ffermion5d}), such as
$q\bar q U\bar U$, are much more suppressed, typically by the light
quark masses.

Finally, we note that the spectrum of KK fermions includes states that typically have masses not very different
from the fourth-generation zero-modes'. Some of these should have
very different signals compared to a standard fourth-generation, given
their exotic quantum numbers~\cite{servant}. 
In general, a very detailed study of all these signals, and the corresponding backgrounds, 
must be carried out in order to assess the reach of the LHC in this model. We leave this for future 
work~\cite{adsewpt}.

\section{Conclusions and Outlook}
\label{conc}
We have shown a viable mechanism for the breaking of the electroweak symmetry and the generation of 
fermion masses through the condensation of a fourth generation. In the context of a 5D theory in 
a slice of AdS$_5$, the super-critical interactions of the 
fourth-generation 
zero-mode quarks are induced by  the KK excitations of the gluon, as well as by 
bulk higher-dimensional operators. These are strong due 
to the localization of the fourth-generation zero-modes close to the IR brane. 
The condensation of the fourth-generation quarks leads to electroweak symmetry breaking and 
results in a heavy Higgs, with a mass $m_h\simeq 900~$GeV, for a KK mass of about $2.5~$TeV. 
The unitarization of SM amplitudes is achieved partially by this heavy Higgs, and partially by the 
presence of the KK gauge bosons. The condensing quarks, the zero-mode of the fourth-generation 
up-quark sector, acquires a dynamical mass of about $m_U\simeq (600-700)$GeV for the same value 
of the KK mass. Larger values of the KK gauge masses result in lighter Higgs and dynamical 
fermion masses. However, as the KK mass is increased the theory becomes more fine-tuned. 

Fermion masses for the lighter three generations, as well as the non-condensing 
fourth-generation fermions, are generated by higher-dimensional bulk operators 
suppressed by the Planck mass. After dimensional reduction, these result in 
four-fermion interactions amongst zero-mode fermions. The ones involving two condensing quarks
will give rise to mass terms for the remaining two fermions. 
This results in the necessary Yukawa textures and the observed fermion masses and mixings. 
Thus, in this model the mechanism of electroweak symmetry breaking
requires flavor violation in the bulk, and is intimately related to
the fermion masses. This is to be contrasted with the proposal of
Ref.~\cite{delocal}, where fermions are de-localized as a way to evade
electroweak constraints; as well as with the one of
Ref.~\cite{giminads}, where there is a flavor symmetry in the 5D bulk.

Regarding electroweak constraints, and 
as is the case with all bulk Randall-Sundrum models where the fermion localization naturally 
explains the fermion mass hierarchy, this model contains a tree-level contribution to the 
$S$ parameter. In addition, the presence of a heavy Higgs results in a positive shift of the 
$S$ parameter at one loop; and the fourth-generation zero-modes induce one loop contributions
to both $S$ and  $T$.  
A full study of the electroweak precision constraints on the model, including the 
contributions from KK modes as well as effects coming from strong coupling, 
is left for a separate publication~\cite{adsewpt}. 
However, we can already conclude that the loop contributions to $S$ are not the decisive factor given 
the presence of a tree-level contribution. In other words, the situation of Randall-Sundrum 
bulk models is not made significantly worse by the presence of a bulk fourth-generation.

The phenomenology of the model at the LHC involves the discovery of a strongly coupled
heavy fourth generation, the signal for a heavy Higgs, typically associated with enhanced 
longitudinal gauge boson scattering. To the usual fourth generation production and decay, 
this model adds the presence of high invariant mass production of the fourth generation through 
its strong coupling to the KK gauge bosons, particularly the gluon. These signals combined
would constitute strong  evidence that the condensation of the fourth-generation quarks is the 
origin of electroweak symmetry breaking~\cite{adsewpt}. 
Other possible phenomenological consequences of the model are, among others, 
the modification of the Higgs production cross section and decay widths~\cite{kribs}, 
flavor physics observables and possible effects in neutrino physics and astrophysics. 
All of them deserve further study.

\bigskip

{\bf Acknowledgments:}
We thank Sekhar Chivukula, Chris Hill and Liz Simmons for 
helpful comments and a reading of the manuscript. 
We also acknowledge the support of the State of S\~{a}o Paulo
Research Foundation (FAPESP). G.B. also thanks the Brazilian  National Counsel
for Technological and Scientific Development (CNPq) for partial support.



\end{document}